\documentclass[fleqn,usenatbib]{mnras}
\usepackage[T1]{fontenc}
\usepackage{ae,aecompl}

\usepackage{graphicx}	
\usepackage{amsmath}	
\usepackage{amssymb}	

\newcommand{\ms}{ h^{-1}{\rm M_{\odot}}}

\title[Inner density Slopes of MaNGA Galaxy]{SDSS--IV MaNGA : The Inner Density Slopes of nearby galaxies}

\author[Ran Li et al.]{
Ran Li$^{1,2,3}$\thanks{E-mail: ranl@bao.ac.cn},
Hongyu Li$^{1,2,3}$,
Shi Shao$^4$,
Shengdong Lu$^{1,2}$,
Kai  Zhu$^{1,2}$,
Chunxiang Wang$^{1,2}$,
\newauthor
Liang Gao$^{1,2,3,4}$,
Shude Mao$^{5,1}$,
Aaron A. Dutton$^{6}$,
Junqiang Ge$^{1}$, 
Yunchong Wang$^5$,
\newauthor
Alexie Leauthaud$^{7}$,
Zheng Zheng$^{1}$,
Kevin Bundy$^{7}$, and
\newauthor
Joel R. Brownstein$^{8}$
 \\
$^{1}$National Astronomical Observatories, Chinese Academy of Sciences, 20A Datun Road, Chaoyang District, Beijing 100012, China\\
$^{2}$ Key laboratory for Computational Astrophysics, National Astronomical Observatories, Chinese Academy of Sciences, Beijing, 100012, China\\
$^{3}$ School of Astronomy and space science, University of Chinese Academy of Science\\
$^{4}$ Institute for Computational Cosmology, Department of Physics, University of Durham, South Road, Durham DH1 3LE \\
$^{5}$ Physics Department and Tsinghua Centre for Astrophysics, Tsinghua University, Beijing 100084, China\\
$^{6}$ New York University Abu Dhabi, PO Box 129188, Abu Dhabi, United Arab Emirates\\
$^{7}$ Department of Astronomy and Astrophysics, University of California, Santa Cruz, 1156 High Street, Santa Cruz, CA 95064, US\\
$^{8}$  Department of Physics and Astronomy, University of Utah, 115 S. 1400 E., Salt Lake City, UT 84112, USA \\
}

\pubyear{2015}

\hypersetup{draft}
\begin{document}
\label{firstpage}
\pagerange{\pageref{firstpage}--\pageref{lastpage}}
\maketitle

\begin{abstract}
We derive the mass weighted total density slopes within the effective (half-light) radius, $\gamma'$, for more than 2000 nearby galaxies from the SDSS-IV MaNGA survey using Jeans-anisotropic-models applied to IFU observations. Our galaxies span a wide range of the stellar mass ($10^9$ $M_{\rm \odot}< M_* < 10^{12}$ M$_{\odot}$) and the velocity dispersion (30 km/s $< \sigma_v <$ 300 km/s). We find that for galaxies with velocity dispersion $\sigma_v>100$ km/s, the density slope has a mean value $\langle \gamma^{\prime} \rangle = 2.24$ and a dispersion $\sigma_{\gamma}=0.22$, almost independent of velocity dispersion. A clear turn over in the $\gamma'-\sigma_v$ relation is present at $\sigma\sim 100$ km/s, below which the density slope decreases rapidly with $\sigma_v$. Our analysis shows that a large fraction of dwarf galaxies (below $M_* = 10^{10}$ M$_{\odot}$) have total density slopes shallower than 1, which implies that they may reside in cold dark matter halos with shallow density slopes. We compare our results with that of galaxies in hydrodynamical simulations of EAGLE, Illustris and IllustrisTNG projects, and find all simulations predict shallower density slopes for massive galaxies with high $\sigma_v$. Finally, we explore the dependence of $\gamma'$ on the positions of galaxies in halos, namely centrals vs. satellites, and find that for the same velocity dispersion, the amplitude of $\gamma'$ is higher for satellite galaxies by about 0.1.
\end{abstract}

\begin{keywords}
galaxies: evolution -- galaxies: formation -- galaxies: kinematics and dynamics -- galaxies: structure
\end{keywords}



\section{Introduction}

Various N-body simulations have shown that the density structure of dark matter halos in a universe formed with pure cold dark matter follows a universal NFW form \citep{NFW96, NFW97}. The density profile, $\rho(r)$, of a dark matter halo scales with $r^{-1}$ for the inner part and scales with $r^{-3}$ for the outer part, independent of the mass of the halo.

In the real Universe the inner density profile depends on both the dark matter and baryons. Baryons cool and condense at the center of dark matter halos and form galaxies. The baryons often contribute a significant amount of mass within the half-light radius \citep[e.g.][]{CourteauDutton2015}. In addition, galaxy formation processes can modify the mass structure of dark matter halos. The baryonic matter can form a much more condensed density profile at the center of a dark matter halo, generating an adiabatic contraction effect \citep[e.g.][]{Blumenthal1986, Gnedin2004, Gustafsson2006}. On the other hand, stellar and Active Galactic Nuclei (AGN) feedback can create rapid gas outflows which may induce an expansion effect at the halo center, resulting in a central core in the dark matter density profiles \citep{Dehnen2005, Read2005,  Duffy2010,  Pontzen2012}. Accurate measurement of density profiles of galaxies thus provides a critical constraint on galaxy formation models.

In observations, the inner density profile of galaxies are best measured through stellar kinematics and gravitational lensing effects \citep[e.g.][]{Koopmans2009, Auger2010, Cappellari2013,Sonnefeld2013, Newman2015,  Poci2017, Huang2018}. The Sloan Lens Advanced Camera for Surveys (SLACS) project studied a sample of 73 galaxy-galaxy lensing systems with lens stellar masses $M_*>10^{11}$M$_{\odot}$ \citep{Koopmans2009, Auger2010} and found that the mean of power-law density slope of the total mass within the Einstein radius to be $\gamma_{\rm tot}=2.078$, very close to a singular-isothermal-sphere profile\footnote{Note that throughout this paper we adopt the convention that density slopes are positive, i.e., $\rho \propto r^{-\gamma}$.}. The dark matter and baryon profiles have very different density slopes, so the close to the isothermal value implies that both baryons and dark matter contribute non-negligible fractions to the mass within the effective radii of massive elliptical galaxies. This is sometimes referred to as the ``bulge-halo conspiracy" \citep{DuttonTreu2014}.

Compared with rare strong lensing systems, dynamical modeling methods can be applied to much larger samples of galaxies. With recent developments in integral field unit surveys, e.g. ATLAS$^{\rm 3D}$ \citep{Cappellari2011}, CALIFA \citep{Sanchez2012}, MASSIVE \citep{Ma2014}, SAMI \citep {Bryant2015} and SDSS-IV MaNGA \citep{Bundy2015},  accurate mass density profiles for large statistical samples of galaxies are becoming available. \citet{Cappellari2013} constructed anisotropic dynamical models for 260 early-type galaxies in the ATLAS$^{\rm 3D}$ sample. Follow-up analysis shows that the inner total density slopes of these early type galaxies have a mean value of $\gamma_{\rm tot}'=2.193$, with a scatter of $\sigma_{\gamma'}=0.174$, which is slightly steeper than the SLACS strong lensing results.  The total density slope is constant for galaxies with effective velocity dispersion $\sigma_v>126$ km/s, and becomes shallower gradually with  $\sigma_v$ for galaxies with $\sigma_v<126$ km/s. For massive clusters, recent observations combining stellar dynamics and gravitational lensing effects show that the total inner density slope drops gradually to 1.7 \citep{Newman2015,Auger2010}.

In this paper, we make use of IFU observations from the MaNGA survey to study the density slopes for nearby galaxies with a wider range of galaxy masses and morphologies than previous studies. We derive the galaxy density profiles using mass models of galaxies in MaNGA DR14 \citep{Abolfathi2018}, which are derived by \citet{LHY2018} with Jeans anisotropic model\citep[JAM,][]{Cappellari2008}, and investigate the relation between density slopes and the galaxy stellar mass, the stellar surface density, the velocity dispersion and other properties.

The structure of the paper is as follows: In Section 2, we outline the MaNGA data we use, in Section 3, we briefly introduce our dynamical modelling technique, and in Section 4, we present our main results, and finally in Section 5, we summarize our results and present a short discussion.

\section{MaNGA galaxies}

The galaxy sample of this project comprises 2778 galaxies from MaNGA survey in SDSS-IV DR14 \citep{Abolfathi2018}, covering a stellar mass range of $10^{9} < M_* < 10^{12}$ M$_{\odot}$, including both early type and late type galaxies. Throughout this work, we define early type galaxies as those with Sersic index $n_{\rm sc} > 2.5$, and the remainders as late type galaxies. Sersic indices are taken from the NASA-Sloan Atlas, NSA catalogue\footnote{Downloaded from https://academic.oup.com/mnras/article-abstract/476/2/1765/4848297 by National Astronomical Observatory user on 26 June 2018}). From the DR14 sample, we exclude merging galaxies (or close galaxy pairs), galaxies with a significant disturbance, and galaxies with strong bars or strong spiral arms. We also exclude galaxies of low data quality, defined as having less than 100 Voronoi with S/N>10 bins. In total, we have 2110 galaxies in our final sample, with 952 early type galaxies and 1158 late type galaxies.

We use the IFU spectra extracted with the official data reduction pipeline \citep[DRP][]{law2016} and the kinematic
data extracted with the official data analysis pipeline \citep{westfall2019}. The kinematic information are obtained by fitting absorption lines using the pPXF software \citep{Cappellari2004, Cappellari2017}.

In this paper, we derive the stellar mass of each galaxy by multiplying their total luminosity in SDSS r-band by their mean stellar mass-to-light ratio within the effective radius, $M_{*}^{\rm SPS}/L^r$, derived by \citet[][see online table]{LHY2018}, in which the mass-to-light ratio is obtained by fitting spectra with stellar population templates using pPXF software \citep{Cappellari2004} with the MILES-based \citep{Sanchez2006} SPS models of \citet{Vazdekis2010} and the \citet{Salpeter1955} initial mass function. Before spectrum fitting, the data cubes are Voronoi binned to S/N=10 \citep{Cappellari2003}. In Appendix B, we show that for a given galaxy, the exact choice of the Voronoi bin number wont' affect the results of the JAM model.

We refer readers to \citet{LHY2018} for details of stellar mass calculation. For MaNGA related information, readers are referred to the following papers: SDSS-IV technical summary\citep{Blanton2017}, MaNGA instrumentation \citep{Drory2015}, observing strategy \citep{Law2015}, sample design \citep{Wake2017}, spectrophotometric calibration \citep{Smee2013, Yan2016a}, survey execution and initial data quality \citep{Yan2016b}, and a SDSS telescope introduction can be found in \citet{Gunn2006} .

\section{Dynamical analysis}

Our fiducial mass models of MaNGA galaxies are built by the method described in \citet{LHY2018} with the multi-Gaussian Expansion \citep[MGE][]{Emsellem1994, Cappellari2002} technique, in conjunction with the Jeans Anisotropic Modelling \citep[JAM][]{Cappellari2008}. MGE decomposes the r-band light distribution of the galaxy with a set of Gaussians, providing an analytical description of the observed surface brightness. The stellar mass distribution is assumed to follow the light distribution of the galaxy in the r-band with a mass-to-light ratio, $\Gamma$, independent of the galaxy radius.  The mass distribution of dark matter halos are assumed to follow gNFW profiles, where density $\rho_{\rm DM}$ can be written as:
\begin{equation}
\label{eq:gnfw}
\rho_{\rm DM}=\rho_s\left(\frac{r}{R_s}\right)^{-\gamma}\left(\frac{1}{2}+\frac{r}{2R_s}\right)^{\gamma-3} \,,
\end{equation}
\noindent where $\rho_s$ is the characteristic density, and $R_s$ is the scale radius. These two parameters are treated as free parameters during fitting.

JAM is then performed to compute the  second moment of the velocity distribution based on the total mass model. Running JAM within an MCMC framework \citep[emcee][]{ForemanMackey2013}, the best-fit mass model that matches the observed second velocity moment map is obtained. We refer the readers to \citep{LHY2018} for details of the mass model constructing. Previous simulations have shown that our method can provide a robust description for total mass profile of galaxies \citep{LHY2016}.

In our fiducial model, we assume light traces mass strictly, i.e. the mass-to-light ratio for the stellar component is a constant, independent of the radius. In the real universe, this assumption is not strictly true \citep[e.g.][]{Poci2017,LHY2018, Sonnenfeld2018, Wasserman2018}. Specifically, in \citet{LHY2018}, we derived the average mass-to-light ratio gradient $\Delta\log{(\Gamma)}/\Delta\log{(R/R_{\rm e})}$ between $R_{\rm e}/8$ and $R_{\rm e}$ using SPS with a Salpeter IMF for MaNGA galaxies in SDSS DR14, and found that the mass-to-light ratio gradient varies from $\sim$-0.2 for massive galaxies to $\sim$0.1 for low mass galaxies. In a recent research by \citet{Sonnenfeld2018} using gravitational lensing data, researchers also derived a mass-to-light ratio of -0.24 for early type galaxies in the SLACS sample.

If the mass-to-light ratio of a MaNGA galaxy is not constant, our best-fit dark matter density profile can be biased, but we still expect that the recovery of the total density profile will not be affected. A validation test is shown in the Appendix, where we assume that the total density profile follows a double power-law model with no prior knowledge for stellar mass distribution. The resulting slopes of the total density profile agree well with that derived from our fiducial model. In this paper, we always show the results using total mass density profile, thus our conclusions are not affected by the mass-to-light ratio gradient.

\section{results}

\subsection{Inner density slopes}


\begin{figure}
\includegraphics[width=0.4\textwidth]{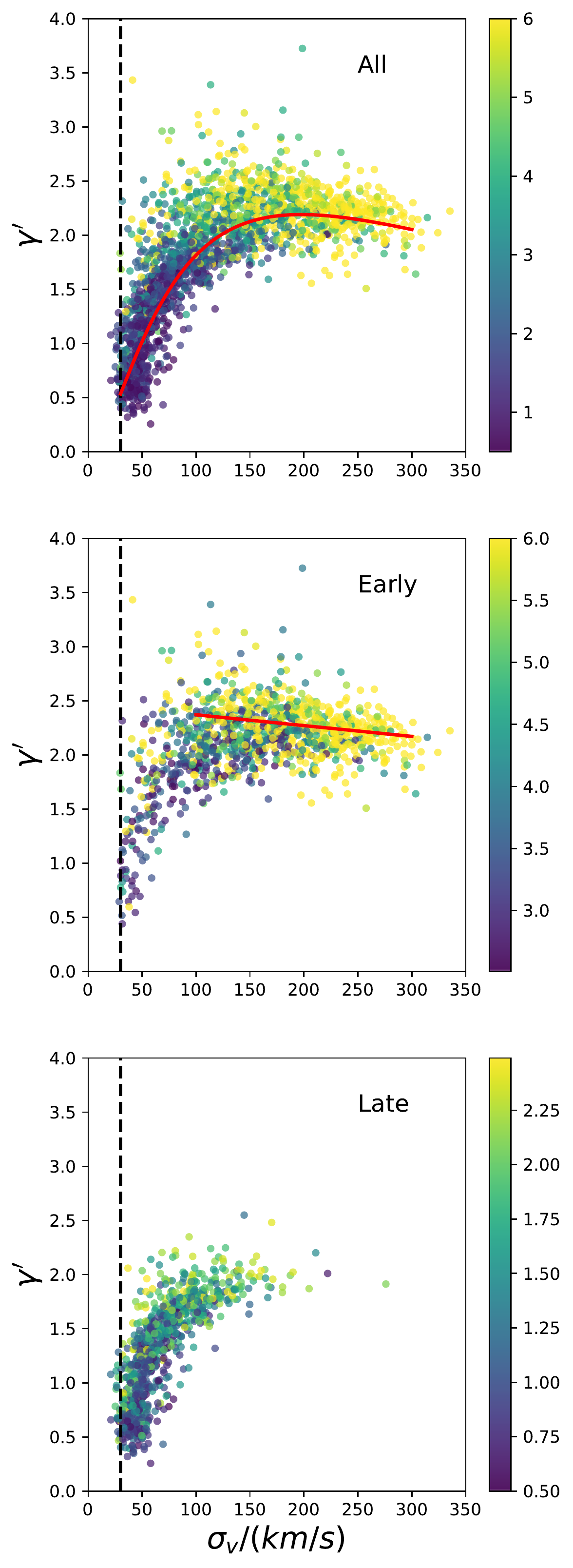}
    \caption{The figure shows $\gamma'$ as a function of $\sigma_v$. The color of the data points show the Sersic index of the light profile, $n_{\rm sc}$. The top panel shows the results for all samples, the middle panel shows the results for early type galaxies ($n_{\rm sc}>=2.5$), and the bottom panel shows the results of late type galaxies ($n_{\rm sc}<2.5$). The red solid line in the top panel shows a best-fit model of Eq.\ref{eq:fit}.  In the middle panel, the red solid line shows a linear fit for data points with $\sigma_v>100$ km/s, $\gamma'=2.47-0.1 \frac{\sigma_v}{100 km/s}$. In each panel, we plot a vertical line at 35 km/s.
     }
    \label{fig:sigma_gm}
\end{figure}

Following \citet{DuttonTreu2014}, we define the mass weighted density slope within the effective radius $R_e$ as
\begin{equation}
\gamma' \equiv -\frac{1}{M(R_e)}\int_0^{R_e} 4\pi r^2 \rho(r) \frac{d\log{\rho}}{d\log{r}}dr= 3 - \frac{4\pi R_e^3\rho(R_e)}{M(R_e)} \,,
\end{equation}
where $\rho(r)$ is total mass density of the galaxy and $M(R)$ is the total mass enclosed in a sphere with radius $R$.

In the top panel of Fig. \ref{fig:sigma_gm}, we present $\gamma'$ for MaNGA galaxies as a function of $\sigma_v$, the luminosity weighted velocity dispersion within $R_e$.  The color of data points show the Sersic index of the light profiles. For galaxies with $\sigma_v<100$ km/s, the density slope increases rapidly from 0.5 to 2, while for galaxies with $\sigma_v \geq 100$ km/s the inner density slopes stay roughly constant at a value of 2.24, slightly steeper than the isothermal slope (2). We find that the $\gamma'-\sigma_v$ relation can be described with the following formula
\begin{equation}
\gamma'= A_0\frac{ (\sigma_v/\sigma_0)^{\alpha}}{(1+\sigma_v/\sigma_0)^{\beta-\alpha}} \,,
\label{eq:fit}
\end{equation}
with $\{A_0,\sigma_0,\alpha,\beta\}=\{19.83,116.62,2.1,5.44\}$.

In the middle and bottom panels of Fig. \ref{fig:sigma_gm}, we divide our galaxies into early type and late type according to their Sersic index and show their $\gamma'-\sigma_v$ relations, respectively. For early type galaxies ($n_{\rm sc} \geq 2.5$) the inner density slope is almost a constant for galaxies with $\sigma_v>100$ km/s. The mean value of the inner total density slope is $\langle \gamma^{\prime} \rangle = 2.24$, with a dispersion $\sigma_{\gamma}=0.22$, agreeing well with previous observations from lensing and stellar dynamics. For those early type galaxies with low velocity dispersion, the total density slope increases rapidly with $\sigma_v$. For late type galaxies, the total $\gamma'$ also increases with $\sigma_v$, the behavior of which is similar to early type galaxies of the same $\sigma_v$ range.

Note that, we don't put any cut on the velocity dispersion for our sample selection. \citet{westfall2019} estimates that the standard deviation of uncertainties on stellar velocity dispersion is $20-30\%$ for a Voronoi bin with S/N >10 and intrinsic stellar velocity dispersion of 35 km/s. In each panel of Fig.\ref{fig:sigma_gm}, we plot a vertical line at 35 km/s. Almost all our galaxies have $\sigma_v> 35$ km/s.


\begin{figure}
\includegraphics[width=0.45\textwidth]{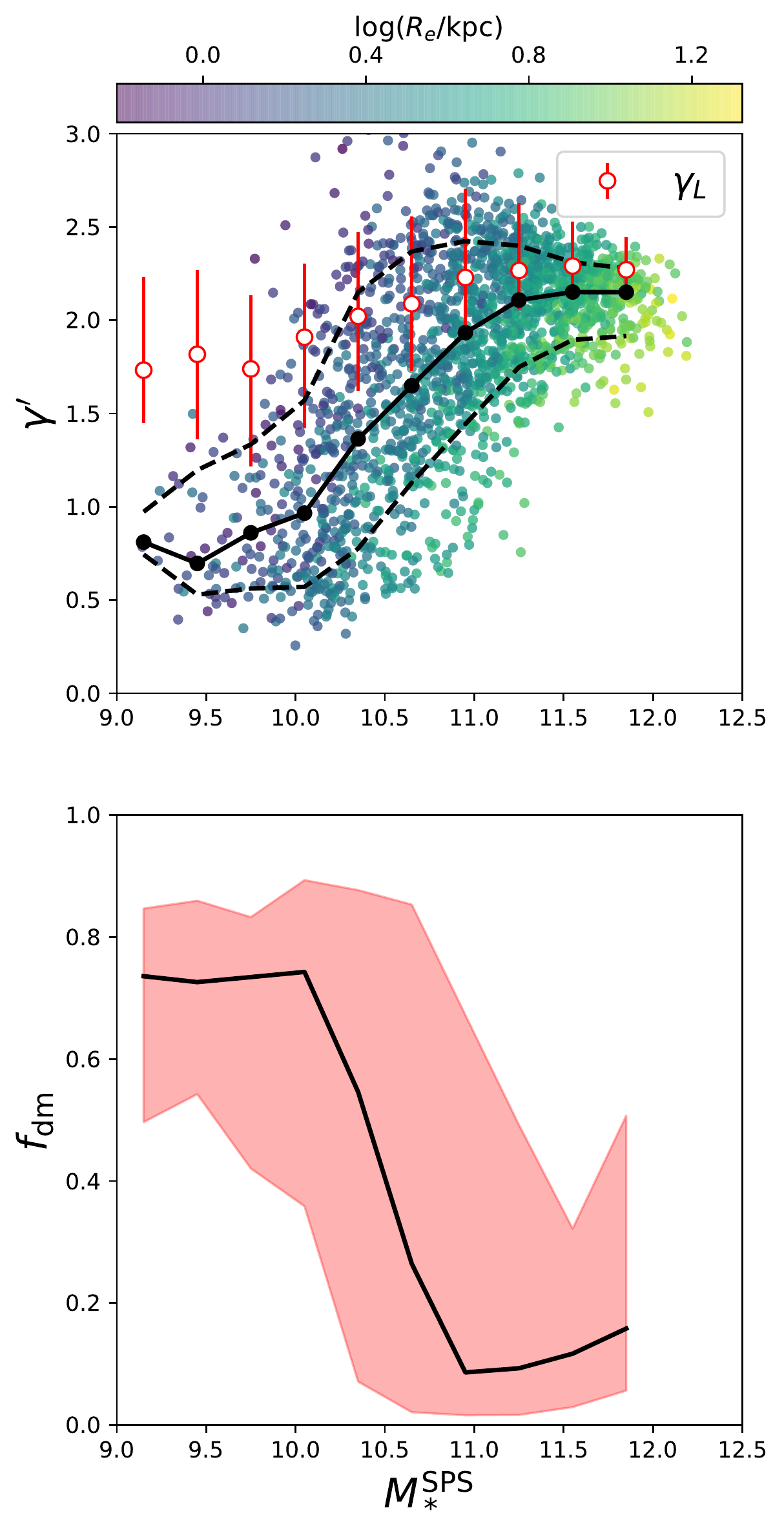}
    \caption{In the upper panel, we show $\gamma'$ as a function of galaxies' stellar mass, $M_*^{\rm SPS}$. The colors of scatter points show the effective radius of these galaxies. The black solid circles represent the mean of the relation and the dashed lines show the region that encloses 70\% sample. For reference, we show the flux weighted r-band brightness slope, $\gamma'_{L}$ as a function of $M_*^{\rm SPS}$ using red empty circles, with error bars showing the region enclosing 70\% of sample.  In the lower panel, we show the $f_{dm}-M_*^{\rm SPS}$ relation with black solid line. The red shadow show the region that encloses 70\% of the sample. }
    \label{fig:mass_gm_re}
\end{figure}

In Fig. \ref{fig:mass_gm_re}, we show the relation between $\gamma'$ and the stellar mass $M_*^{\rm SPS}$. In the upper panel, we plot $\gamma'$ vs. $M_*^{\rm SPS}$ for each MaNGA galaxy using scatter points. The colors of points represent the effective radius of each galaxy in kpc. The black solid circles represent the  mean  of  the  relation  and  the two dashed lines enclose 70\% of the galaxies at each mass. For reference, we show the flux weighted r-band brightness slope of the stars, $\gamma'_L$, as a function of stellar mass using red empty circles.  The average slope of the luminous matter is always steeper than the average total mass slope, implying that the average total slope is an upper limit to the average density slope of the dark matter within the effective radius. In the lower panel of the figure, we plot $f_{\rm dm}$, the dark matter to total mass fraction with $R_e$, which is derived from JAM model. The density slope grows almost linearly with $M_*^{\rm SPS}$ in the range of $[10^{10},10^{11}]$ $M_{\odot}$, while for the most massive galaxies, $\gamma'$ goes flat with stellar mass with a mean of $\sim 2.15$.

One can find that the galaxies of larger sizes tend to have lower $\gamma'$ for the same stellar mass. Larger effective radii affect the total density slope in two ways: a more extended distribution of stellar mass which results in a shallower stellar density slope;  more dark matter included into the calculation of $\gamma'$, because the dark matter usually takes larger fraction of total mass at the outer part of a galaxy. Since the dark matter density slopes are much shallower than those of stellar mass, the latter effect also drives the total density slope to a lower value.

For galaxies more massive than $10^{11} M_{\odot}$, the amplitude of mean $\gamma'-M_*^{\rm SPS}$ flattens. From the lower panel of Fig. \ref{fig:mass_gm_re}, we find these galaxies contain little dark matter, thus their total density slopes are almost identical as their stellar density slopes, which is also roughly constant at this mass range (red empty circles in the upper panel of Fig. \ref{fig:mass_gm_re}).


Interestingly, we find that the mean total density slopes for galaxies less massive than $10^{10}M_{\odot}$ are below 1.0, while their luminosity profile slopes still possess a mean value of $\sim 1.7$ and the JAM modelling shows these galaxies are dominated by dark matter within the effective radius. A direct implication of these results is that many galaxies in this mass range reside in dark matter halos with shallow inner density slopes even shallower than 1. In the context of CDM, dissipationless simulations predict steeper slopes than 1, thus the shallow slopes we find for MaNGA dwarf galaxies implies their dark matter halos have expanded.

The $M_*/L$ gradient can also contribute to the shallow total density profile. However, \citet{LHY2018} has shown that the $M_*/L$ gradient of MaNGA galaxies with $M_*^{\rm SPS}< 10$ is $\sim 0.1$ ( positive sign means the central value is higher). Thus the observed shallow total density profile are not likely to be explained purely by the gradient of $M_*/L$.

\begin{figure*}
    \centering
    \includegraphics[width=0.8\textwidth]{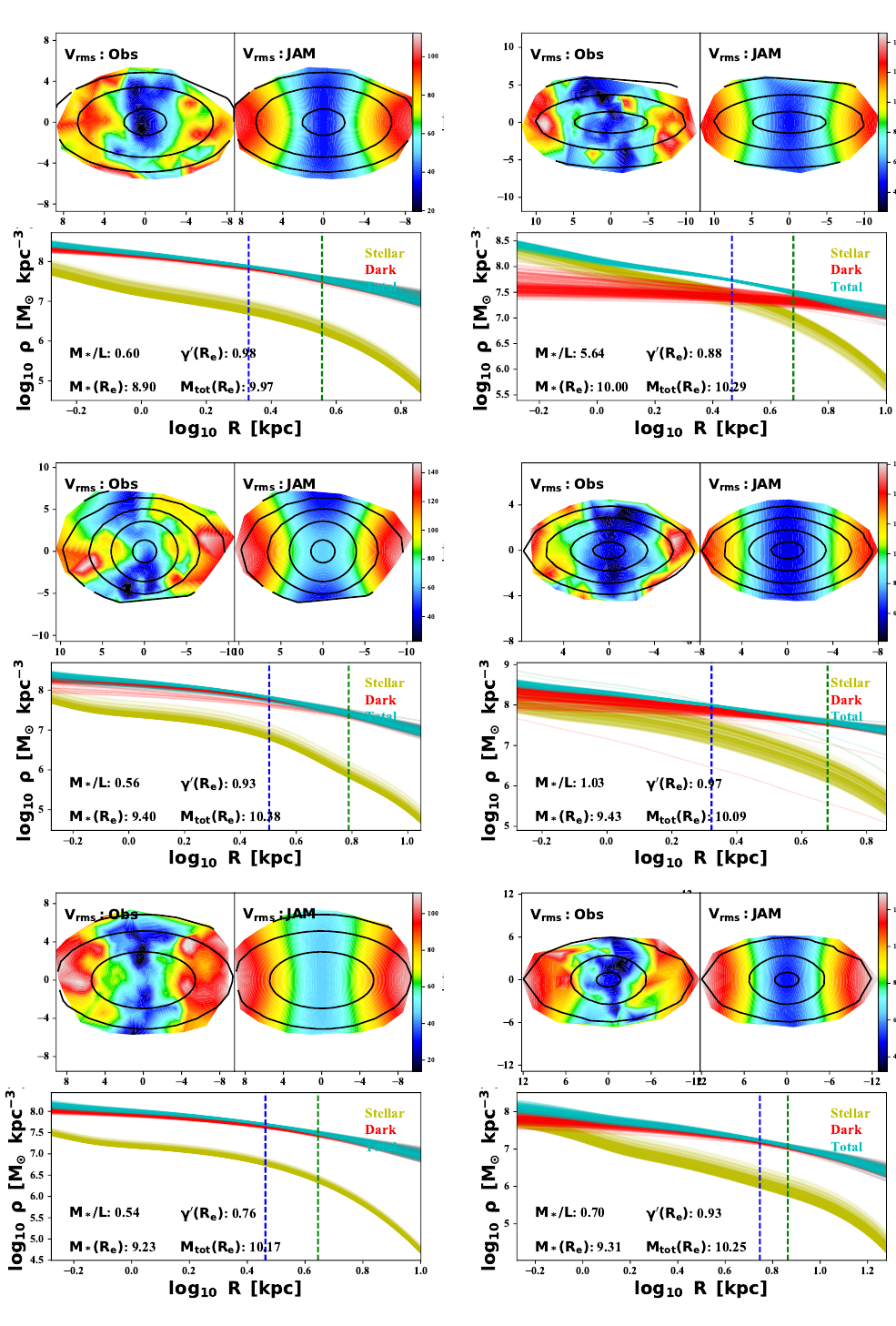}
    \caption{In this figure, we show 6 examples of galaxies with total density slopes $\gamma'<1$. In the upper panel of each subplot, we show the observed velocity field, and the velocity field reconstructed with the JAM. The unit for the axes is arcsec. In the lower panel, we show the density profiles derived from the JAM model. For each galaxy, we randomly select 200 points from posterior distribution of model parameters and plot the density profiles calculated with these parameter combinations using solid thin lines. The yellow, red and blue lines show profiles of stellar, dark matter and total mass, respectively. The vertical lines show the position of $R_e$ and $2R_e$. In each subplot, we also mark the quantities derived using JAM, including: stellar mass-to-light ratio,  the total density slope, the total mass enclosed within $R_e$, and the total stellar mass enclosed within $R_e$ }
    \label{fig:examples}
\end{figure*}

One caveat of the results is that the galaxies are all dwarf galaxies of relatively small physical size and luminosity, thus may suffer larger systematics in JAM modelling. However, we find the density slopes of the dwarf galaxies not correlating with their apparent sizes. Dwarf galaxies with relatively larger apparent size can still have shallow density slopes. In Fig.\ref{fig:examples}, we show 6 examples of galaxies with total density slopes $\gamma'<1$. In the upper panel of each subplot, we show the distribution of observed  root-mean-squared velocity $v_{\rm rms}=\sqrt{(v_{\rm los}^2 + \sigma_{\rm los}^2)}$, and the $v_{\rm rms}$ reconstructed with the JAM. In the lower panels, we show the density profiles derived from the JAM model. For each galaxy, we randomly select 200 points from posterior distribution of model parameters and plot the density profiles calculated with these parameter combinations using solid thin lines. The yellow, red and blue lines show profiles of stellar, dark matter and total mass, respectively. For these examples, we can find they all have steep gradients of of $v_{\rm rms}$, with high $v_{\rm rms}$ at $R_e$ and low $v_{\rm rms}$ at inner part, which requires flat total mass density slopes within $R_e$.

\begin{figure}
\includegraphics[width=0.5\textwidth]{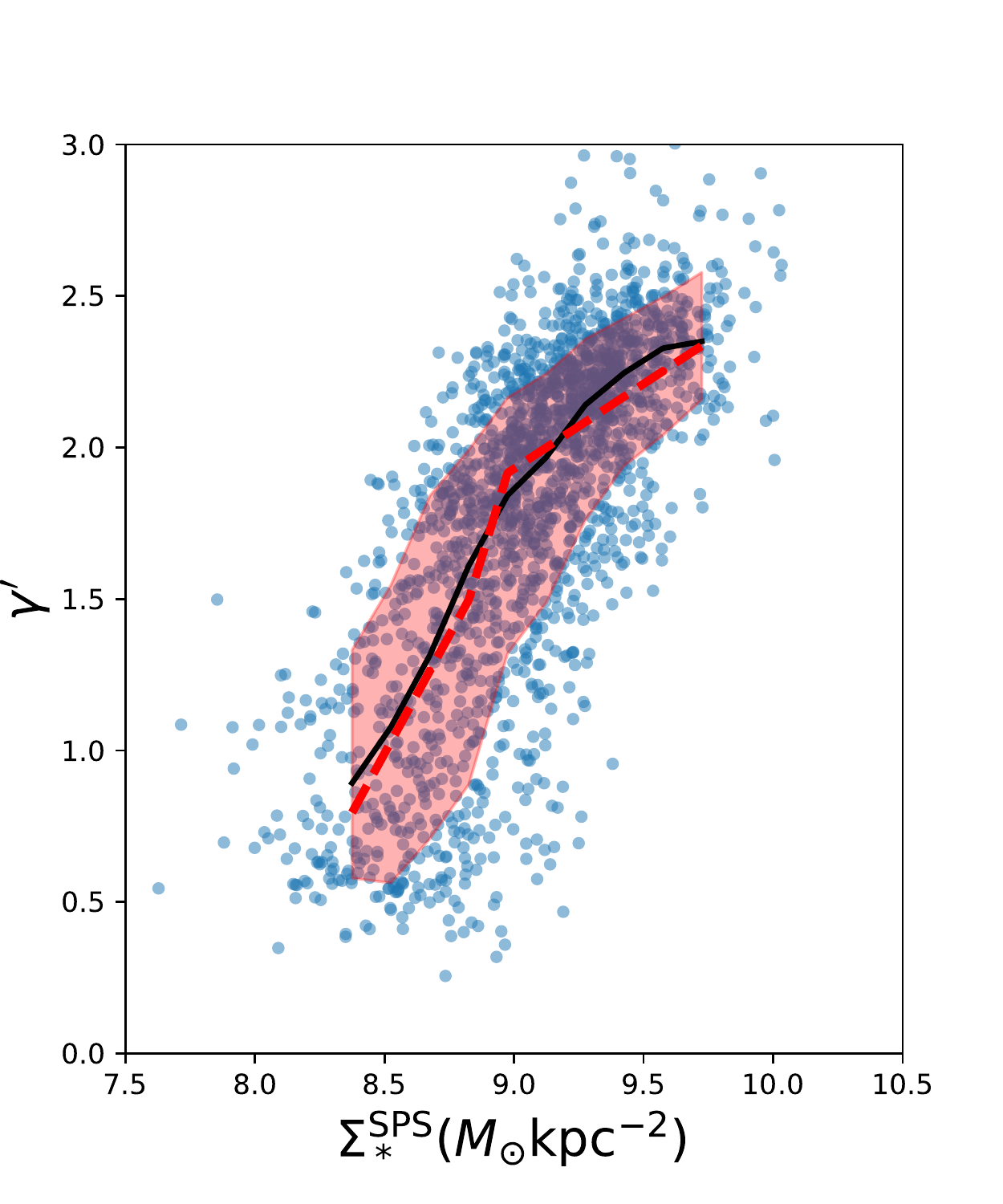}
    \caption{The figure shows the relation between $\gamma'$ and stellar surface density $\Sigma_*^{\rm SPS}$. Blue points show the data of of each galaxy. The red shadow shows the region enclosing 70\% data points and the black line shows the mean of the relations. The red dashed line in the right panel shows the best-fit double linear function for $\gamma'-\Sigma_*^{\rm SPS}$. }
    \label{fig:sd_gamma}
\end{figure}

 In Fig. \ref{fig:sd_gamma}, we show the relation between $\gamma'$ and  $\Sigma_*^{\rm SPS}$, the stellar surface density averaged within $R_e$. Blue points show the data of of each galaxy. The red shading shows the region enclosing 70\% data points and the black line shows the mean of the relation. We find the relation between $\gamma'$ and $\Sigma_*^{\rm SPS}$ changes slope at $\sim 10^{9} M_{\odot}/{\rm kpc}^2$, and we can fit the relation with
\begin{equation}
\gamma'= \left\{
\begin{array}{cc}
a_1\log{M} + b_1 & \text{if $M$ < $M_s$} \\
&\\
a_2\log{M} + b_2 & \text{if $M$ > = $M_s$}
\end{array} \right .,
\end{equation}
where $\{M_s,a_1,b_1,a_2,b_2\}=\{8.90,0.56,-3.11,1.56,-12.27\}$.

\begin{figure}
\includegraphics[height=0.8\textheight]{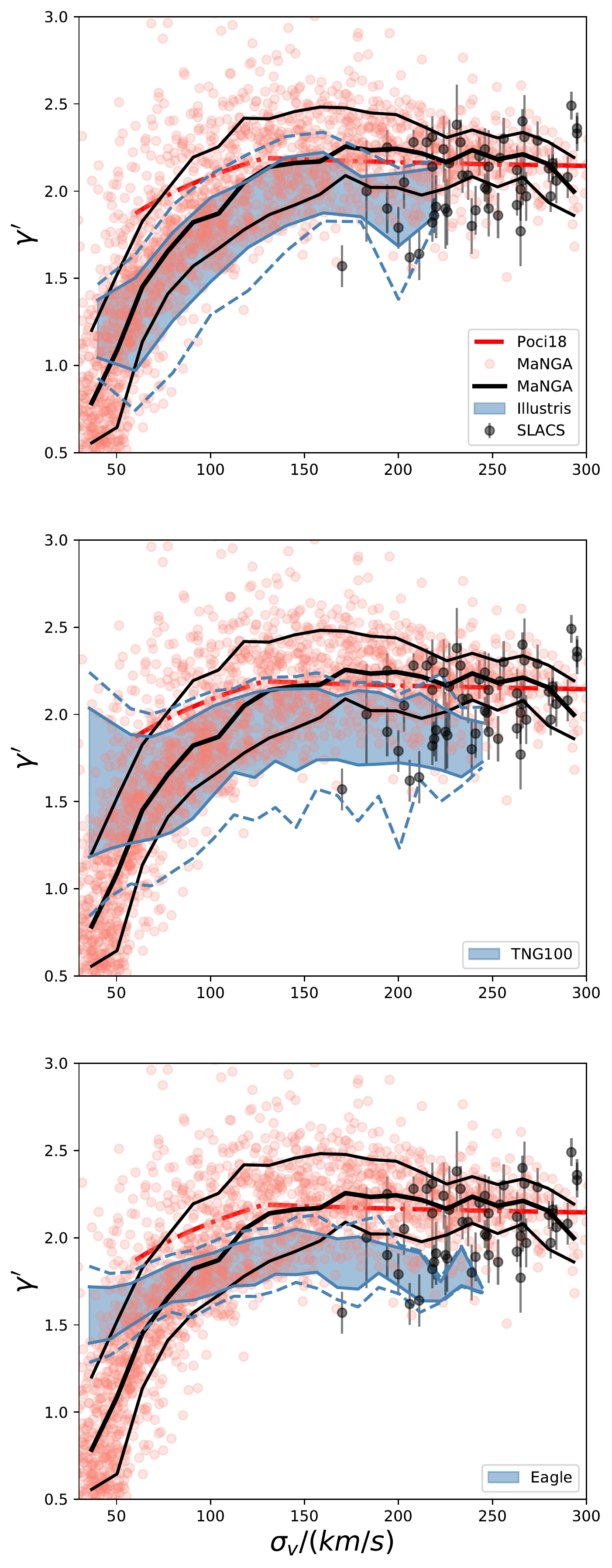}
    \caption{Red data points show observational results in this work. The mean and the region enclosing 70\% red points are shown with black solid lines. The black data points show $\gamma'$ from strong lensing observation from \citep{Auger2010}. And the
    dashed line show the results from \citep{Poci2017}.
    From top to bottom panels, we compare the results with Illustris, TNG100 and EAGLE simulations respectively. The blue shading shows the region enclosing 70\% of the simulated galaxies and the dashed lines show the region enclosing 90\% simulated galaxies.  }
    \label{fig:compare_simulation}
\end{figure}

In Fig. \ref{fig:compare_simulation}, we compare our density slopes with predictions from hydrodynamical simulations: EAGLE \citep{Schaye2015,Crain2015} and Illustris \citep{Genel2014, Vogelsberger2014a,Vogelsberger2014b} and IllustrisTNG project \citep{Marinacci2018,Naiman2018,Nelson2018, Pillepich2018a, Springel2018, Nelson2019}. For IllustrisTNG simulations, we only use the simulation run in cubic volume of 100 Mpc side length (TNG100, hereafter). In these simulations we select all galaxies with stellar mass $M_*>10^9M_{\odot}$, and calculate their total $\gamma'$ within the 3D radius that equals to their projected half-light radius. To calculate the projected half-light radius, we assume a constant mass-to-light ratio.

In general, all simulations reproduce the total density slope of galaxies reasonably well. However, all of the simulations predicts a mean $\gamma'$ slightly smaller than 2 for the galaxies with high $\sigma_v$ , which is shallower than the observed value of 2.24. At the low $\sigma_v$ end, the Illustris simulation can reproduce the shallow density slopes in observations, while the total density slopes predicted by EAGLE simulation are significantly higher than the observed values. For the TNG100 simulation, the density slopes spread a wide range for low $\sigma_v$ galaxies. The turn over of $\gamma'-\sigma_v$ relation at $\sigma\sim100$ km/s is recovered in the Illustris, while the EAGLE and the TNG100 simulation predict flat $\gamma'-\sigma_v$ relations.

In this figure, we also plot the total density slopes from strong lensing of the SLACS \citep{Auger2010} and the JAM modelling results for ATLAS$^{\rm 3D}$ early type galaxies \citep{Poci2017}.  Our results are consistent with previous strong lensing observations and ATLAS$^{\rm 3D}$ results for early type galaxies with $\sigma_v>100$ km/s. For galaxies with lower velocity dispersion, our analysis gives a lower mean $\gamma'$ than \citet{Poci2017}, which mainly because our samples include late type galaxies which have lower total density slopes.

\begin{figure*}
\includegraphics[width=\textwidth]{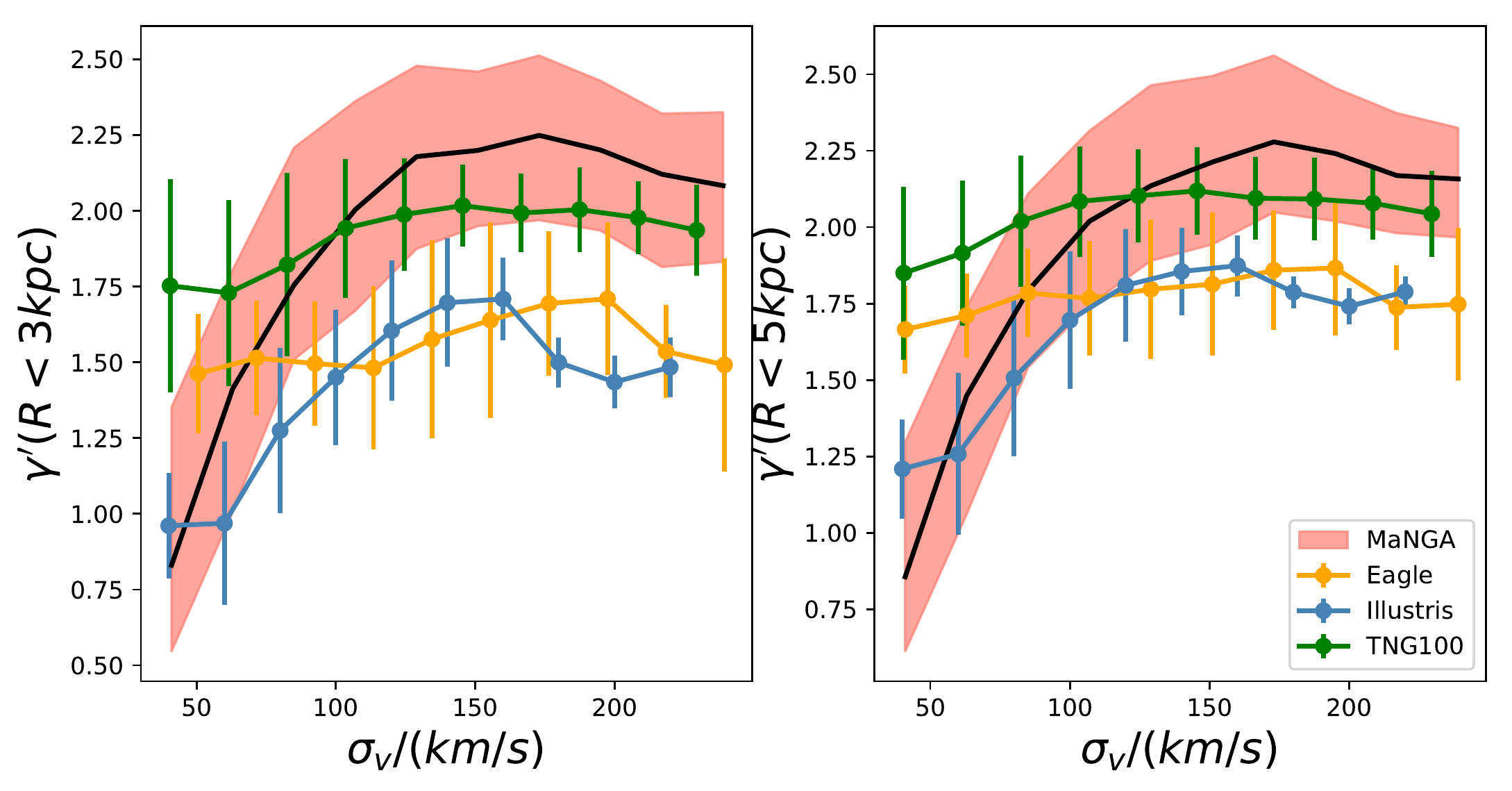}
    \caption{The figure shows the mass weighted total density slopes within 3 kpc (left) and 5 kpc (right), respectively. The shadings show the regions enclosing 70\% observed galaxies and the black solid lines show the mean. The blue, green, and yellow lines show the mean $\gamma-\sigma_v$ relation of Illustris, TNG100 and EAGLE simulations and the error bars show the region enclosing 70\% of simulated galaxies.  }
    \label{fig:fix_radius}
\end{figure*}

Since the hydrodynamical simulations may not reproduce perfectly the observed galaxy mass-size relation, the definition of half-light radius may not be referring to the same physical radius in observations and simulations. In Fig. \ref{fig:fix_radius}, we further compare the total mass weighted density slope between simulated galaxies and our MaNGA sample within fixed radius of 3 kpc and 5 kpc. Again, for fixed radius of 3 kpc and 5 kpc, the simulations predict smaller $\gamma'$ for galaxies with higher $\sigma_v$. The differences are more significant for Illustris and EAGLE simulations than that for TNG100 simulation.

\subsection{central vs satellite}

In this project, we match the galaxies to the group catalog derived by  \citet{Yang2007}
(hereafter SDSSGC\footnote{http://gax.shao.ac.cn/data/Group.html}).
SDSSGC is constructed with the spectroscopic galaxy sample
of SDSS DR7 \citep{Abazajian2009} using an adaptive halo-based
group finder. Each galaxy in SDSS DR7 is assigned to a group,
and a halo mass is estimated for each group. We assume the
central galaxy of each group is the one that has largest stellar
mass.

For each galaxy in our MaNGA sample, we find its counter part in
the SDSSGC. Therefore, we can obtain its host halo mass, and we
can tell whether it is a central galaxy or a satellite galaxy.

\begin{figure}
\includegraphics[width=0.5\textwidth]{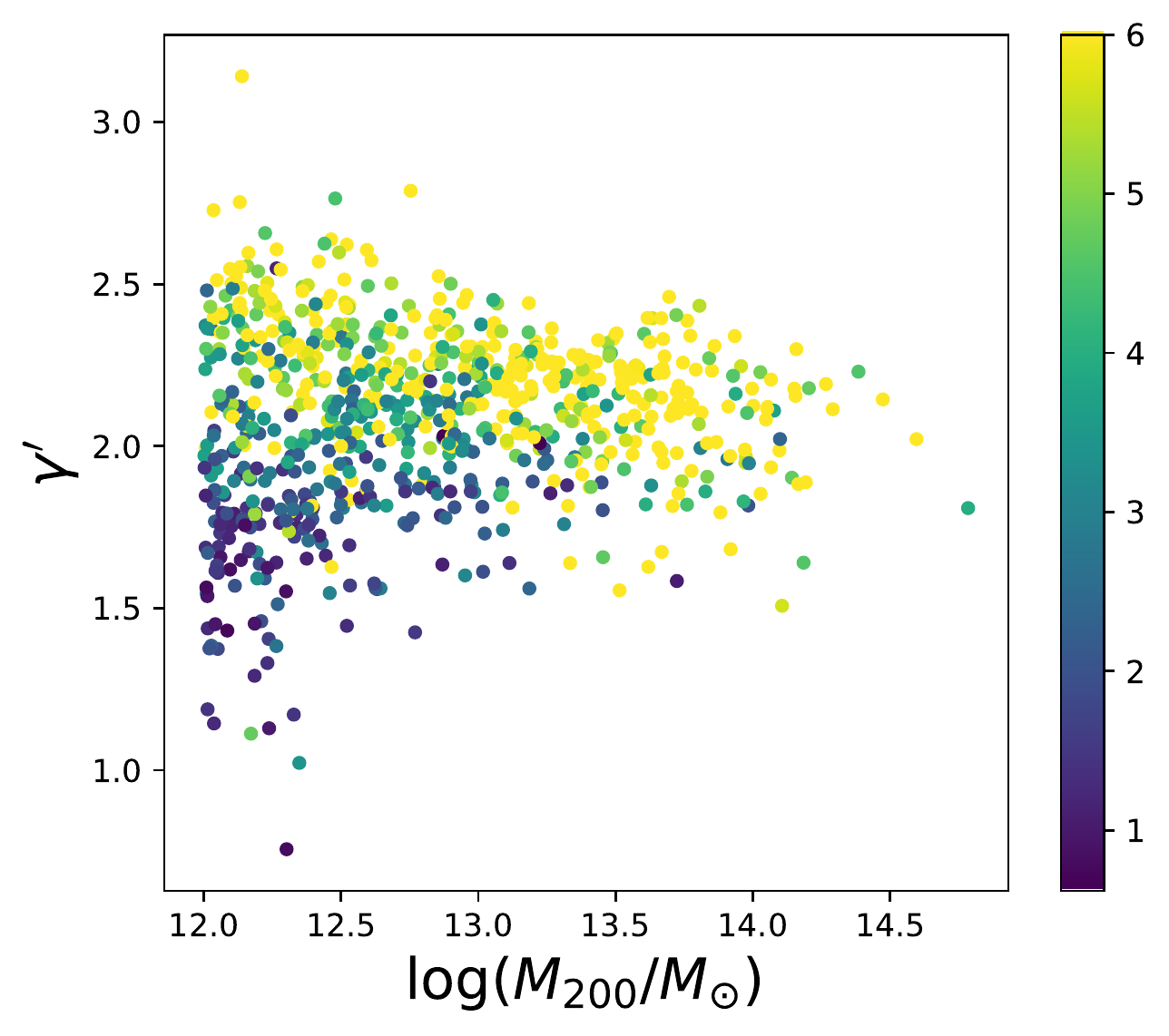}
    \caption{The figure shows the total mass density slope of
    the MaNGA galaxies as a function of $M_{\rm 200}$
    of their groups. We only plot the results for central galaxies. The color of data points show the Sersic index of galaxies.}
    \label{fig:gm_halo}
\end{figure}

In Fig. \ref{fig:gm_halo}, we select only the central galaxies
from the MaNGA sample, and plot their total mass density slopes
as a function of $M_{\rm 200}$ of the groups. For groups with
mass larger than $10^{13}$ $\ms$, their central galaxies are
mostly early type whose total mass density slope has a mean
value 2.08 and a scatter of 0.27. For the lower mass groups, the
central galaxy could be either early type or late type, and
their total slopes scatter in a broad range from 1.5 to
2.75.

\begin{figure}
\includegraphics[width=0.5\textwidth]{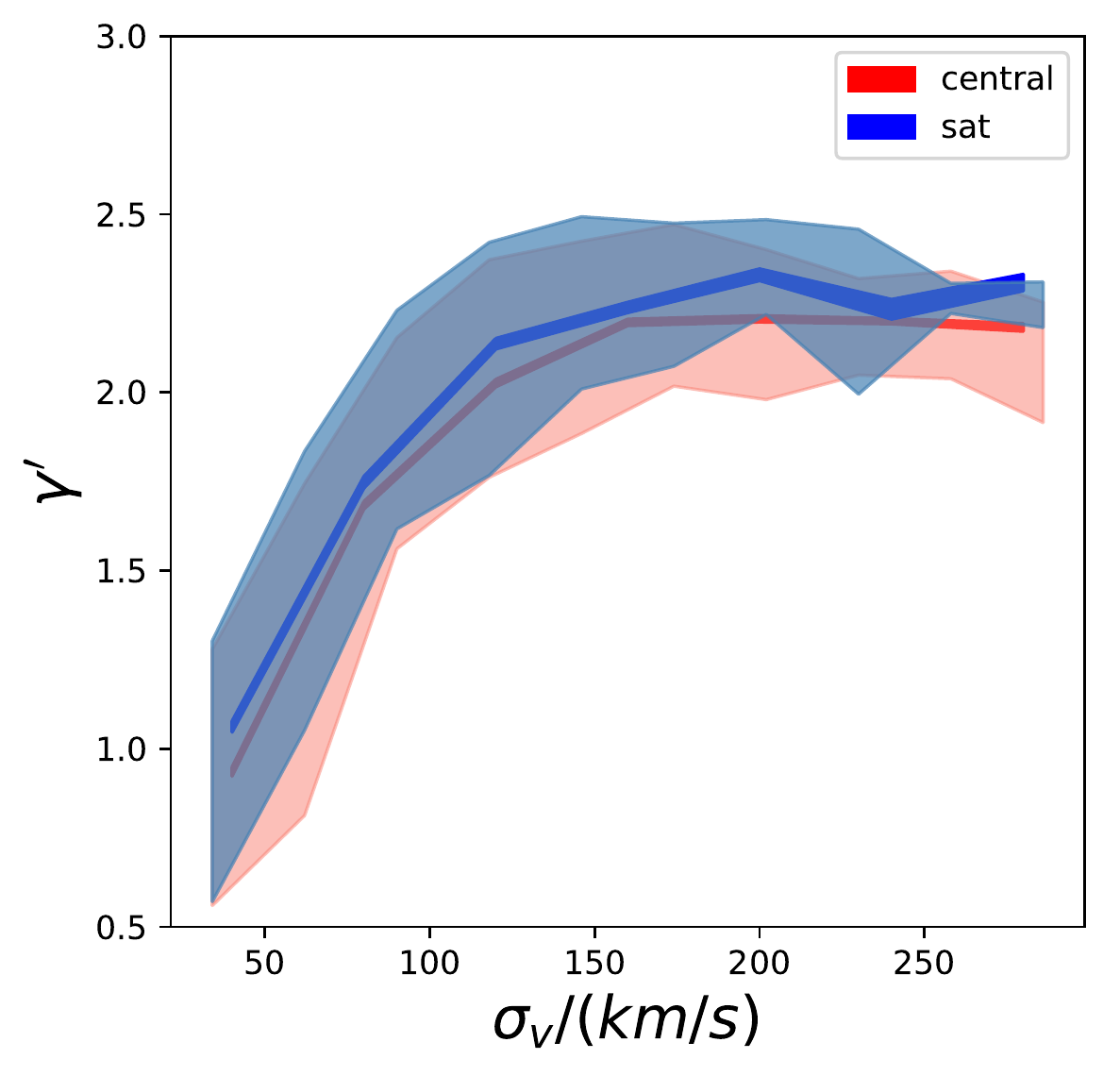}
    \caption{In Fig. \ref{fig:cen_sat}, we show mean $\gamma'-\sigma_v$ relation for central (red) and satellite (blue) galaxies, respectively. The line width show the uncertainties of the mean $\gamma'-\sigma_v$ relation. The shading show the regions enclosing 70\% of central (red) and satellite (blue) galaxies. }
    \label{fig:cen_sat}
\end{figure}

In Fig. \ref{fig:cen_sat}, we show the total mass slope $\gamma'$ for
central galaxies and satellite galaxies using red and blue
lines, respectively. The overall dependence of $\gamma'$ on
$\sigma_v$ is similar for centrals and satellites. For galaxy with the
similar $\sigma_v$, the amplitude of the mean $\gamma'$ for satellites is
higher than that of central by $\sim 0.1$, which maybe due to the fact
that on average satellite galaxies form earlier in denser universe and
have more concentrated stellar profiles.

\section{Summary and discussion}

In this paper, we investigate the inner density profile of more than 2000 nearby galaxies from the  SDSS-IV MaNGA survey.  Our sample of galaxies is the largest of its kind, spanning 3 decades of stellar mass, and includes both early type galaxies and late type galaxies.  Using IFU observations we derive $\gamma'$, the mass weighted total mass density slope within the effective radius, for these galaxies from the mass model built with the JAM method.

For galaxies with $\sigma_v>100$ km/s,  their mean density slopes hold almost a constant value of $\gamma'= 2.24$, decrease slowly with increasing of velocity dispersion, while for galaxies with $\sigma_v<100$ km/s, the density slopes decrease rapidly with decreasing velocity dispersion.

We also investigate the relation between $\gamma'$ and the stellar mass derived using stellar population synthesis models. We find the density slopes increase linearly with $\log{M_*^{\rm SPS}}$ in the mass range of $[10^{10},10^{11}] M_{\odot}$, but is roughly constant for more massive galaxies at a value of about 2.15.  For these most massive galaxies, we find they have little dark matter within $R_e$, and their total density slopes are almost the same as their stellar mass density slopes.

For the same stellar mass, galaxies of larger sizes tend to have lower total density slopes. We find $\gamma'$ increases with average stellar density within Re, with slopes of 0.56 for galaxies with $\log{(\Sigma_{*}^{\rm SPS}/(M_{\odot} {\rm kpc^2})) } <8.9$ and 0.156 for galaxies  $\log{(\Sigma_{*}^{\rm SPS}/(M_{\odot} {\rm kpc^2}))}>8.9$.

Interestingly, we find that many of the galaxies less massive than $10^{10} M_{\odot}$ have small total density slopes, even shallower than $1$. In the context of cold dark matter, which predicts slopes of 1 in the absence of baryons, our results imply expansion of the dark matter halo. A recent paper by \citet{Benitez-Llambay2018} shows that shallow dark matter density profile at inner part of the halo closely relates to the star formation threshold density. A star formation threshold density higher than 1 cm$^{-3}$  allows the gas to pile up at the center of halo, so that a energetic feedback afterwards can efficiently blow out the gas and flatten the inner dark matter profile \citep[also see][]{Bose2018, Dutton2018}. A future detailed comparison of dwarf galaxy mass structures between observations and simulations may shed light on the subgrid feedback process in galaxy formation.

We compare our results with hydrodynamical simulations, EAGLE, Illustris and TNG100. In all simulations, the total density $\gamma'$ for galaxies with $\sigma_v>150 $ km/s are slightly below 2, shallower than our observed value of 2.24.

Finally, we explore the density slope dependence on their positions in groups/clusters, namely whether a galaxies is central galaxy or satellite galaxy. We divide our MaNGA galaxies into centrals and satellites using the SDSS group catalog by \citet{Yang2007}. For early type central galaxies, the density slopes are about $\gamma'\sim2.08$, and decreases slowly with increasing  $M_{\rm 200}$. For late type central galaxies, the density slope scatters in a wide range from 1.5-2.75. We also find that the amplitude of the $\gamma'-\sigma_v$ relation for satellite galaxies is higher than that of centrals by 0.1.

\section*{Acknowledgements}
We acknowledge National Natural Science Foundation of China (Nos. 11773032, 11390372, 1133303, 11821303, 11761131004), and National Key Program for Science and Technology Research and Development of China (2017YFB0203300). RL is supported by NAOC Nebula Talents Program and Newton mobility award. This work is partly supported by the National Key Basic Research and Development Program of China (No. 2018YFA0404501 to SM)
SS is supported by grant ERC-StG-716532-PUNCA and STFC [grant number ST/L00075X/1, ST/P000541/1]

We performed our computer runs on the Zen high performance computer cluster of the National
Astronomical Observatories, Chinese Academy of Sciences (NAOC), and the Venus server
at Tsinghua University. This research made use of Marvin, a core Python package and web framework for MaNGA data, developed by Brian Cherinka, Jos{\'e} S{\'a}nchez-Gallego, and Brett Andrews. (MaNGA Collaboration, 2017).

Funding for the Sloan Digital Sky Survey IV has been provided by
the Alfred P. Sloan Foundation, the U.S. Department of Energy Office of
Science, and the Participating Institutions. SDSS-IV acknowledges
support and resources from the Center for High-Performance Computing at
the University of Utah. The SDSS web site is www.sdss.org.

SDSS-IV is managed by the Astrophysical Research Consortium for the
Participating Institutions of the SDSS Collaboration including the
Brazilian Participation Group, the Carnegie Institution for Science,
Carnegie Mellon University, the Chilean Participation Group, the French Participation Group,
Harvard-Smithsonian Center for Astrophysics,
Instituto de Astrof\'isica de Canarias, The Johns Hopkins University,
Kavli Institute for the Physics and Mathematics of the Universe (IPMU) /
University of Tokyo, Lawrence Berkeley National Laboratory,
Leibniz Institut f\"ur Astrophysik Potsdam (AIP),
Max-Planck-Institut f\"ur Astronomie (MPIA Heidelberg),
Max-Planck-Institut f\"ur Astrophysik (MPA Garching),
Max-Planck-Institut f\"ur Extraterrestrische Physik (MPE),
National Astronomical Observatories of China, New Mexico State University,
New York University, University of Notre Dame,
Observat\'ario Nacional / MCTI, The Ohio State University,
Pennsylvania State University, Shanghai Astronomical Observatory,
United Kingdom Participation Group,
Universidad Nacional Aut\'onoma de M\'exico, University of Arizona,
University of Colorado Boulder, University of Oxford, University of Portsmouth,
University of Utah, University of Virginia, University of Washington, University of Wisconsin,
Vanderbilt University, and Yale University.




\bibliographystyle{mnras}
\bibliography{ref_manga} 




\appendix

\section{Dependence on the choice of parameterization}

In our fiducial model, we assume the stellar mass distribution follows the light distribution and the dark matter halo follows a NFW profile. To explore whether our results depend on the choice of mass model, we perform JAM with an axisymmetric double power-law model
\begin{equation}
\rho_{\rm tot}(l)=\rho_s(\frac{l}{l_s})^{\gamma}(\frac{1}{2}-\frac{1}{2}\frac{l}{l_s})^{-\gamma-3} \,,
\end{equation}
where the ellipsoid radius $l=\sqrt{R^2+z^2/q^2}$, where $z$-axis is the symmetric axis, and $R$ is the transverse radius; Free parameters q, $l_s$, and $\rho_s$ are the intrinsic axis ratio, scale radius, and density at the scale radius, respectively. In this model, we don't split density into light and dark when calculating the gravitational potential. For light density tracer, we still use the same MGEs as our fiducial model.

\begin{figure*}
\includegraphics[width=\textwidth]{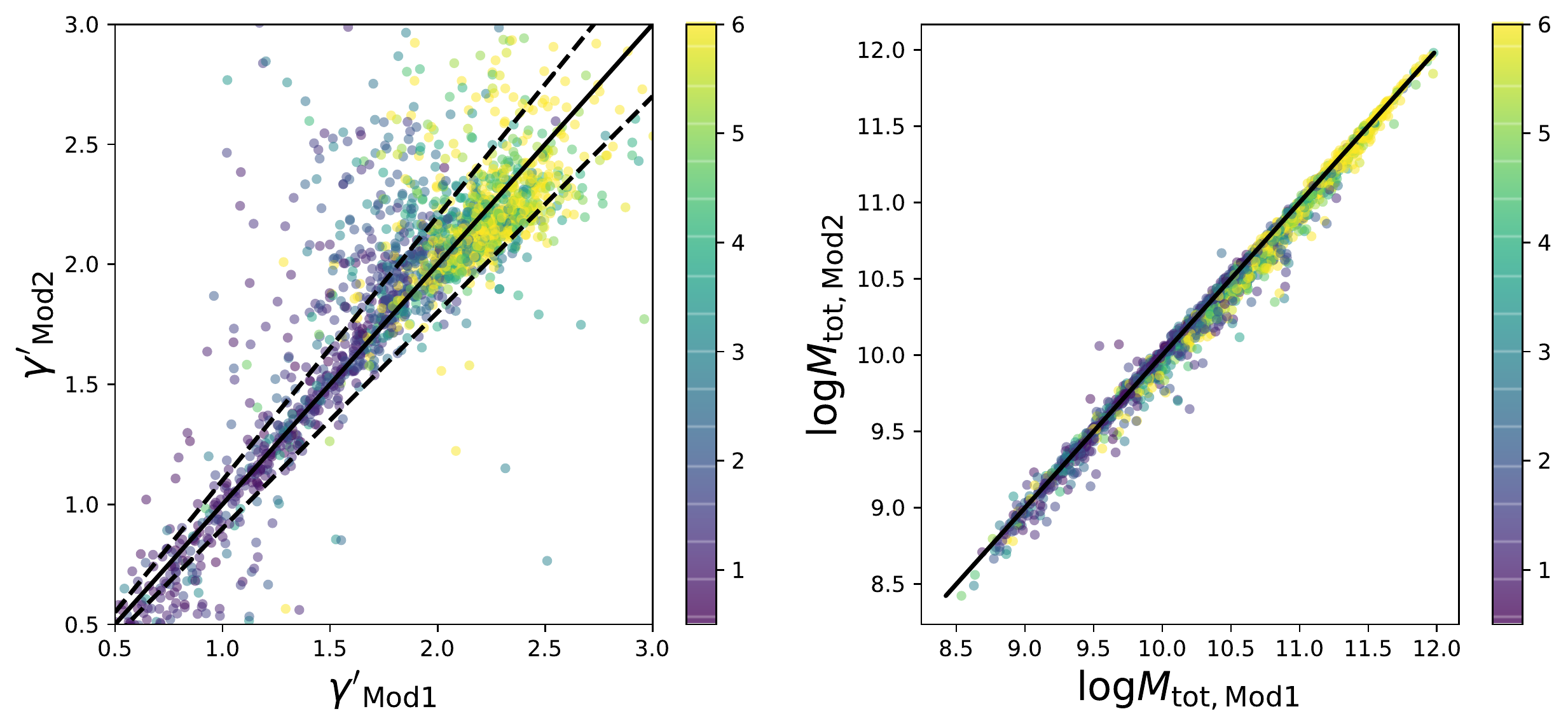}
    \caption{In the left panel, we compare the recovered $\gamma'$ from our fiducial model (Mod1) and
    axisymmetric double power-law model (Mod2). The solid line shows y=x, and the 10\% scattering region is shown by two dashed lines. The color represent the Sersic index of the galaxies. In the right panel, we compare the total mass within the effective radius from the two models.}
    \label{fig:model_compare}
\end{figure*}

 In Fig. \ref{fig:model_compare}, we compare the mass weighted total density slope and the total mass within $R_e$. One can find that the recovered values of total mass within the effective radius from two models agree well with each other. Two different mass models also produce consistent total density slope. Although the dispersion increases slightly for galaxies with $\gamma'>2$, but the results for the majority of the galaxies still agree between the two models. Thus, we assert that our conclusions in this paper are not affected by the detailed choice of parameterization.

\section{Dependence on the number of Voronoi bins}
\begin{figure*}
    \includegraphics[width=\textwidth]{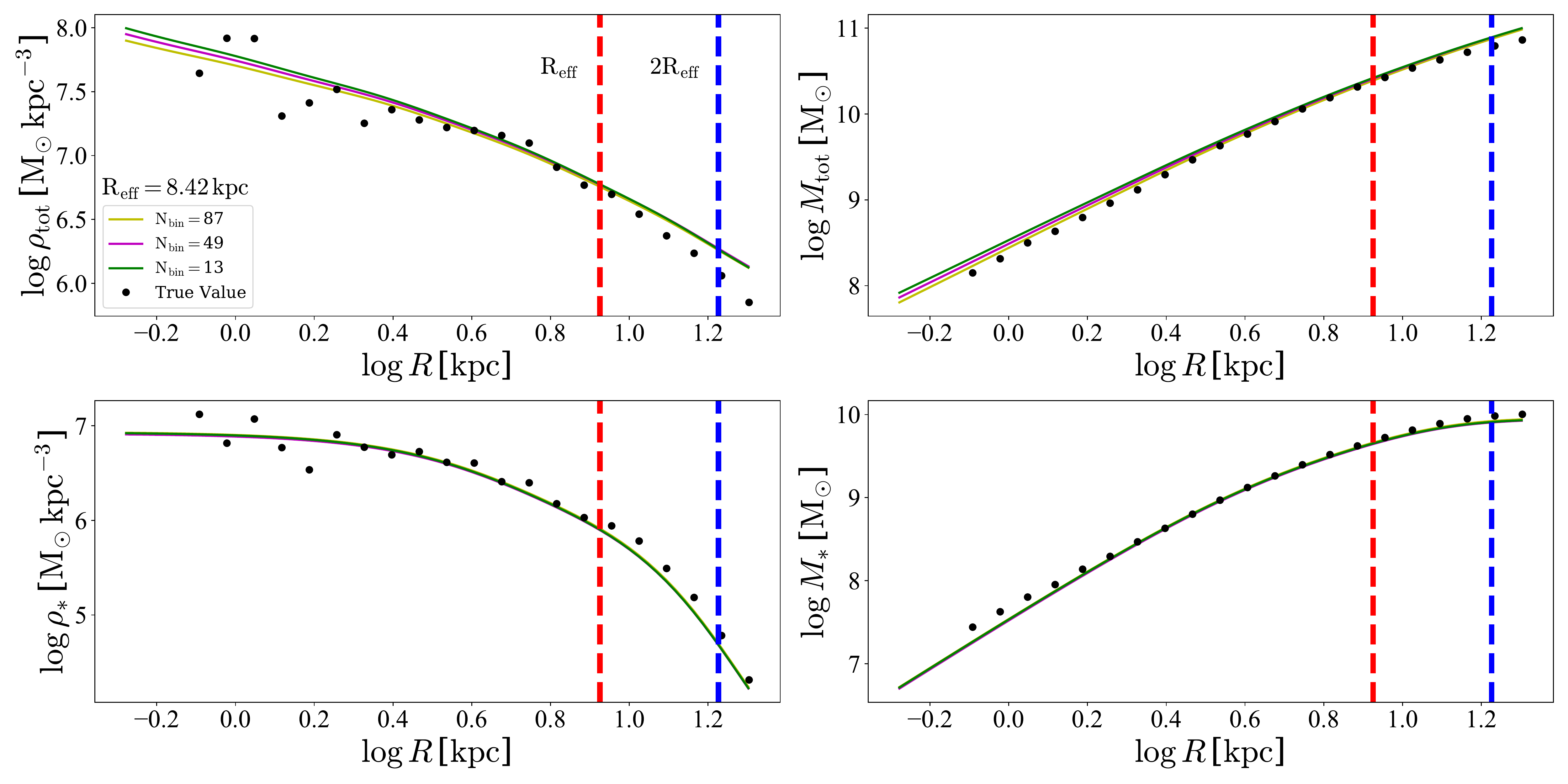}
    \caption{The figure shows the recovery of the total density profile (top left), the stellar density profile (bottom left), the total mass enclosed within R (top right), and the total stellar mass within R (bottom right) for a simulated galaxy from Illustris simulation. The vertical lines show the position of $R_e$ and $2R_e$. In each figure, different lines show the results using different number of Voronoi bins.}
    \label{fig:bin}
\end{figure*}

Before deriving the kinematic information from spectrum fitting, the MaNGA data cubes are Voronoi binned \citep{Cappellari2003} to an S/N=10, which may introduce a spatial smoothing effect to the kinematic data.  To test the dependence of results on the number of Voronoi bins, we have tried use a binned criteria of S/N=60, and find the results doesn't change. To further explore the effect of binning, we run our JAM analysis on a simulated galaxy from Illustris simulation. We derive the brightness and velocity fields of the galaxy within $R_e$ on regular grids of $2\times2$ kpc$^2$, comparable to the physical size covered by a single MaNGA fibre. Then, we Voronoi bin the velocity grids with three different S/N criteria. The numbers of bins within $R_e$ are 13, 49 and 87.  For each binned data set, we perform the JAM  analysis. In Fig.\ref{fig:bin}, we show the recovery of the total density profile (top left), the stellar density profile (bottom left), the total mass enclosed within R (top right), and the total stellar mass within R (bottom right) for different binning conditions. We find the recovery of the mass model of a galaxy are not affected by the choice of number of Voronoi bins.

\bsp	
\label{lastpage}
\end{document}